\documentclass[12pt,a4paper]{article}
\usepackage[T2A]{fontenc}
\usepackage[cp1251]{inputenc}
\usepackage[english]{babel}
\usepackage{amssymb}
\usepackage{amsmath}
\usepackage{graphicx}
\usepackage[small,sc,center]{titlesec}
\usepackage{indentfirst}
\usepackage[labelsep=period]{caption}
\usepackage{caption}
\newcommand{\msun}{\,$M_{\odot}$}

\newcommand{\rsun}{\,$R_{\odot}$}

\newcommand{\ergs}{\,erg\,s$^{-1}$}

\newcommand{\gcm}{\,g\,\,cm$^{-1}$}

\newcommand{\kms}{\,km\,s$^{-1}$}

\newcommand{\fei}{Fe\,{\sc i}}

\begin{document}
	
\begin{center}
\textbf{\large Supernova in lost common envelope and SN~2006gy}
	
	\vskip 5mm
\copyright\quad
2025 \quad N. N. Chugai \footnote{email: nchugai@inasan.ru}\\
\textit{Institute of astronomy, Russian Academy of sciences, Moscow} \\
Submitted  20.05.2025
\end{center}

{\em keywords:\/} stars  -- supernovae; stars -- binary stars; 

\noindent
{\em PACS codes:\/} 

\clearpage
 
 \begin{abstract} 
 A mechanism is proposed for synchronizing core-collapse supernova with 
   a recent loss of a red supergiant (RSG) envelope in the common envelope regime.
 A prerequisite for the synchronization is a moderate RSG expansion during final decade.  
  This scenario is based on the phenomenon of preSN~II dense shell 
  formed at the final stage of 10-20 yr as a result of powerfull mass loss.
The energy deposition into the RSG envelope that powers the enormous mass loss rate 
 is able to expand the RSG.
The moderate expansion is sufficient for the close secondary component  
  to plunge into the  common envelope with a subsequent explosion of stripped 
    helium core.   
 Superluminous SN~2006gy is suggested to be the outcome of this scenario.

\end{abstract} 	

\section{Introduction}

Unique supernovae powered by the ejecta interaction with a close very massive circumstellar (CS) 
  envelope ($M_{cs} > 5$\msun), viz. SN~2006gy (Ofek et al. 2007; Smith et al. 2007) 
    and SN~2008iy (Miller et al. 2010; Chugai 2021) raise a question on the 
    mechanism of the massive envelope loss before the supernova explosion.
The most likely possibility is the loss of a common envelope (CE) in a binary system 
  (Chugai \& Danziger 1994; Chugai \& Chevalier 2006; Ofek et al. 2007; Chevalier 2012; Jerkstrand et al. 2020).
 However this poses the next uneasy question, how does supernova know that it should 
   explode soon after the CE loss?
To put it another way, what does synchronize the loss of 
the common envelope with the subsequent supernova explosion?     

Currently three synchronization mechanisms for the supernova and the
 preceding loss of the massive envelope are proposed: (i) neutron star plunge into the CE 
   is accompanied by the neutron star spin-up and field amplification with the subsequent CE loss 
   and magneto-rotational explosion (Barkov \& Komissarov 2011);
  (ii) white dwarf spiral-in results in CE loss and subsequent white dwarf explosion as SN~Ia (Jerkstrand et al. 2020); (iii)  ejection of a massive shell several years prior to the final explosion of the pusational pair-instability SN (PPISN) (Woosley et al. 2007).
 
Here I propose synchronization mechanism that suggests the explosion of the 
 naked core SN (NCSN) soon after the CE loss.
 The interaction of CCSN with a lost CE has been already proposed for 
 SN~2001em (Chugai \& Chevalier 2006) and  SN~2006gy (Ofek et al. 2007). 
 This scenario is described below in more detail with the demonstration,  
  how the frequency of these events is related to the behavior of 
  massive star 10-20 yr prior to the core collapse.
I describe conditions for the CCSN  to interact with the recently lost massive envelope 
 and explore the application of this scenario to puzzling superluminous SN~2006gy.

\section{NCSN after CE loss}

Stars with the initial mass of 9-25\msun\ end up with the supernova 
   explosion as a result of iron core collapse (Woosley et al/ 2002).
The optical display of the explosion can be either SN~II (IIP or IIL), if the star retains the 
   hydrogen envelope, or the naked core supernova (NCSN, i.e. SN~Ibc/SN~IIb), if the star looses the hydrogen envelope as a result of the binary evolution in the CE (Podsiadlowski et al. 1992; Woosley et al. 2002).
The CE loss is a key process for the scenario proposed below for  
 NCSN with close massive CS envelope.   
 
Despite the detailed desription of the binary system in the CE is still lacking, the 
 following statements seem to be rather robust.    
First, if the binary separation ($a$) is equal to the preSN radius ($R_{rsg}$) the 
  the binary system enters the CE.
Second, for the RSG presupernova with the fiducial mass of $\approx15$\msun\ the secondary with the mass $\gtrsim 2$\msun\ is able to remove the CE 
 (Kruckow et al. 2016).
Third, if the binary already is in the CE, the latter will be lost in several orbital periods 
 (Ivanova et al. 2013). 
For the total binary mass of $\sim 15$\msun\ and $a \approx R_{rsg} \approx 800$\rsun\ the orbital period is 
   $\sim 1.8$ yr,  so the CE will be lost in $\sim 5$ yr after the binary entering the CE.

\subsection{NCSN and \"{O}pik distribution}

The \"{O}pik distribution (\"{O}pik 1924 )  for the binary orbital separation 
 $dN/d\lg{a} = const$  is commonly considered as a sensible approximation through the five order separation range $a_2/a_1 = 10^5$ with $a_1= 10$\rsun\  (Popova et al. 1982; Vershchagin et al. 1988; Han et al. 2020). 
To validate the universal feature  of this distribution one needs to  
 check that it is able 
 to recover the observed NCSN/SN~II ratio $\mathcal{R} = N(\mbox{NCSN})/N(\mbox{SNII}) = 0.33$  (Lee et al. 2010). 

A binary system with a separation less than the RSG radius at the helium burning stage 
  ($R_{\mbox{\tiny{He}}}$) forms the CE that will be lost, while the remaining helium core, 
  possibly with traces of hydrogen, will explode as NCSN.
 The fraction of these binaries with respect to massive binaries is $f_1 = 0.2\lg{(R_{\mbox{\tiny{He}}}/a_1)}$.
Binaries with $a > R_{\mbox{\tiny{He}}}$ evolve in two ways.
First, those with the separation less than the RSG radius at the carbon burning stage 
 (note $R_{\mbox{\tiny{C}}} > R_{\mbox{\tiny{He}}}$) form the CE that will be lost and 
 the helium core explosion produce NCSN; the fraction of theses binaries is 
 $f_2 = 0.2\lg{(R_{\mbox{\tiny{C}}}/R_{\mbox{\tiny{He}}})}$. 
Second, massive binaries with $a > R_{\mbox{\tiny{C}}}$ retain the hydrogen envelope and explode as SN~II; their fraction is 
  $f_3 = 0.2\lg{(a_2/R_{\mbox{\tiny{C}}})}$.
The ratio NCSN/SN~II is, therefore, 
 \begin{equation}
   \mathcal{R} = \frac{\phi(f_1 + f_2)}{\phi f_3 + 1 - \phi}\,,
   \label{eq:ratio}
   \end{equation} 
  where $\phi$ is the fraction of massive binaries with a primary in the range $9-25$\msun.
  
For the fiducial preSN mass of 15\msun, the RSG radius at the helium and carbon burning stage 
  is $R_{\mbox{\tiny{He}}} = 500$\rsun\ and $R_{\mbox{\tiny{C}}} = 800$\rsun, respectively 
     (Woosley et al.2002).
 Given $R_{rsg}$ values, expressions for $f_1$, $f_2$, $f_3$ and Equation (\ref{eq:ratio}) one   
  obtains from the
    condition $\mathcal{R} = 0.33$ the required binary fraction $\phi = 0.65$, which is 
    in accord with the fraction of OB-star binaries (Duch\'{e}ne \& Kraus 2013).
We thus confirm that the \"{O}pic distribution reproduces the supernovae ratio 
   NCSN/SN~II for conventional parameters, so this distribution can be used to 
  estimate fractions of CCSN varieties.

The ratio of NCSNe lost the CE at the helium and carbon 
  burnig stages for the fiducial primary of 15\msun\ turns out to be  $\lg{(R_{\mbox{\tiny{He}}}/a_1)}/\lg{(R_{\mbox{\tiny{C}}}/R_{\mbox{\tiny{He}}})}     \approx 8.5$.  	
This estimate shows that most NCSNe loose their hydrogen envelope approximately  
   $t_{\mbox{\tiny{He}}} \sim 2\times10^6$ yr (Woosley et al. 2002) before the explosion and small fraction 
   of NCSNe loose their envelope $ t_{\mbox{\tiny{C}}} \sim 2\times10^3$ yr (Woosley et al. 2002)  before the explosion.
 The NCSN interaction with the lost massive envelope thus occures not earlier than  
 $t_{coll} \sim t_{\mbox{\tiny{C}}}(v_{cs}/v_{sn}) \sim 20$ yr after the explosion, 
 for the supernova velocity  $v_{sn} = 10000$\kms\ and the velocity of the expanding CE $v_{cs} = 100$\kms.
 
A multiband search for the CS interaction of NCSNe at the epochs $\sim 20$ yr after the explosion, particularly in the radio band (Soderberg et al. 2006), is crucial for the verification of the NCSNe scenario.
We, however, focus at 
   the rare events, when NCSN shows a powerful interaction  with a very 
    massive CS envelope shortly ($\sim 1$ month) after the explosion. 
    These events apparently are missing in the conventional scenario for the NCSNe formation, 
     
 \subsection{ NCSN synchronization with CE loss}
 
 Supernovae that start to interact with the massive CS shell soon ($\sim 1$ month) after 
  the explosion are supposed to loose the CE recently, 
  $t_{loss} \approx \mbox{(1 month)}\times(v_{sn}/v_{cs}) \sim 10$ years, before the 
  explosion.
What does connect, at first glance independent events, --  CE loss and subsequent 
  supernova explosion?

The answer is suggested by the ubiquitous presence of a dense confined shell (DCS) among  
 SNe~II  (Khazov et al. 2016; Chugai 2001; Yaron et al. 2017) recovered from early supernova   spectra (1-4 days). 
The DCS outer radius is of $R_{ds}\sim (0.5-1)\times10^{15}$\,cm 
(Chugai 2001; Yaron et al. 2017) 
and the mass is from $\sim 0.004$\msun\ (Yaron et al. 2017) up to $\sim 0.1$\msun\ (Chugai 2001).
With the typical RSG wind velocity $u_w \approx (10-20)$\kms\ 
  the DCS should be formed by the enormous mass loss rate during 
  $t_{ds} = R_{ds}/u_w \sim 10-20$\,yr before the supernova explosion.
The proximity of time scales of DCS formation $t_{ds}$ and the CE loss $t_{loss}$ is 
  striking and signals us that these phenomena, indeed, are closely linked.
  
The mass loss mechanism responsible for the DCS formation is not yet understood. 
It is probably related to the energetic processes at the final stage of nuclear burning.
For example, the gravity waves generated by the vigorous convection could be converted into 
 acoustic waves that deposit their energy in the RSG envelope (Shiode \& Quataert 2014). 
Regardless the specific mechanism, the deposited power should result in the expansion of 
  the RSG envelope. 
It is the presupernova expansion during the final 10-20 yr before the colapse that 
  provides the link between CE loss and the subsequent supernova explosion.
  
Let, at the stage of the DCS formation, the RSG radius increases by $\Delta R = \epsilon R_{rsg}$.
With a finite probability, a close binary separation falls in the range    
  $ R_{rsg} < a <  R_{rsg}(1 + \epsilon)$.
If this is the case, the binary turnes out to be in the CE, which will be lost   
   in several orbital periods 10-20 yr before the helium core explosion.   
To demonstrate the effect,  let us adopt $\epsilon = 0.1$.
In this case the fraction of NCSNe showing a signature of the powerful CS interaction 
 with respect to all NCSNe turns out to be  
  $\lg{(1+\epsilon)}/\lg{(R_{\mbox{\tiny{C}}}/a_1)} \approx 0.02$. 
For $\epsilon = 0.2$ this fraction becomes 0.04. 

Thus, the moderate expansion of the RSG by only 10\% during the final 10-20 years before 
the core collapse can result in the loss of the presupernova hydrogen envelope in the CE regime and the 
 subsequent explosion of NCSN with the powerful CS interaction in 2\% cases of NCSNe.
The specific feature of this phenomenon is the explosion of helium core followed by 
  the ejecta interaction with the lost massive CE at the radius $R_{cs} \sim v_{cs}t_{ds} \sim (3-6)\times 10^{15}(v_{cs}/100\mbox{\kms})$ cm. 
  
 %================================================================
 \begin{figure}
 	\centering
 	\includegraphics[trim= 0 200 0 100,clip, width=\columnwidth]{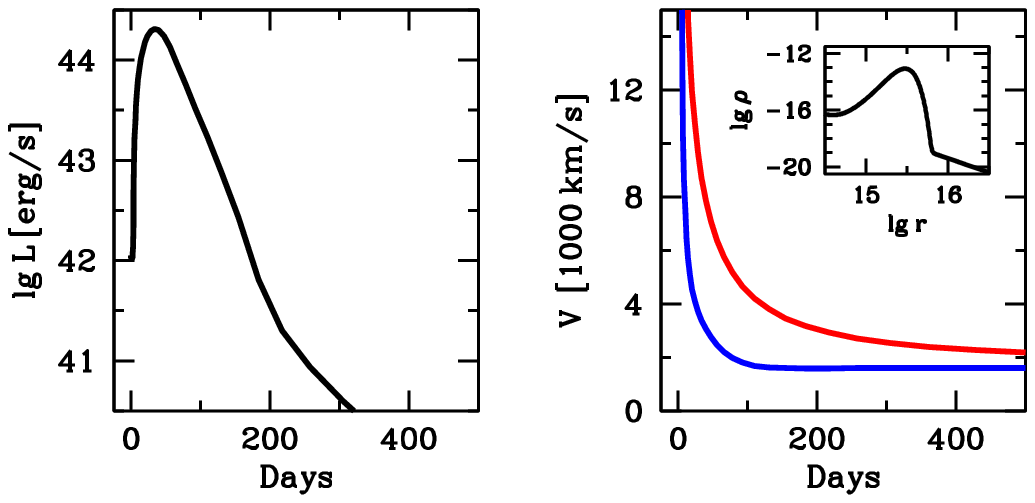}
 	\caption{
 		{\em Left.} Bolometric light curve of the model A (Table 1). 	
 		{\em Right.}  Velocity of the CDS ({\em blue}) and maximum velocity of unperturbed ejecta 
 		({\em red}). 
 		Inset shows CS density distrbution.
 	}
 	\label{fig:rcs3}
 \end{figure}
 %==================================================================
 
 %================================================================
 \begin{figure}
 	\centering
 	\includegraphics[trim=0 200 0 100,clip, width=\columnwidth]{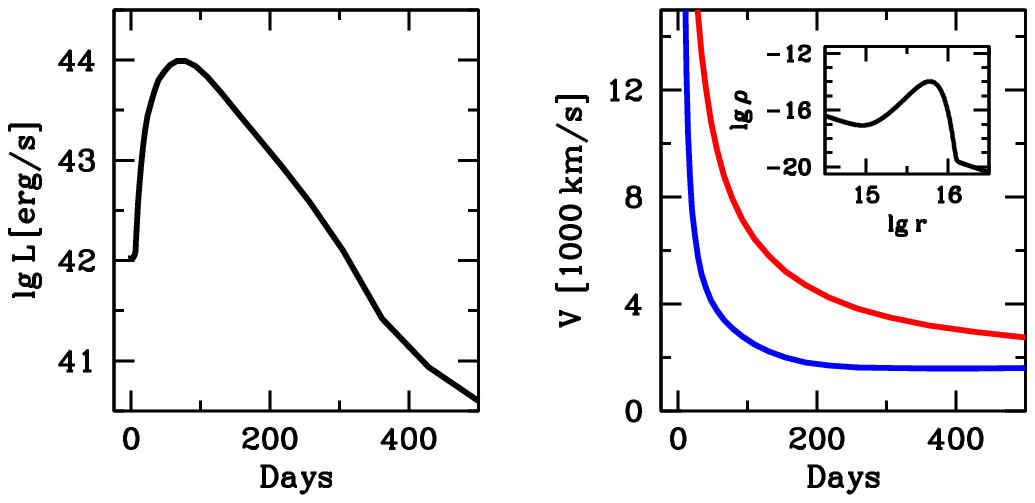}
 	\caption{
 		The same as Figure 1, but for the model B.	
 	}
 	\label{fig:rcs6}
 \end{figure}
 %==================================================================

 \subsection{NCSN after recent CE loss} 
 
 The optical outcome of NCSN soon after the CE loss is considered for  
  the primary initial mass of 14-15\msun\  that ends up with the helium core of 4\msun\ 
     (Woosley et al. 2002).    
The two versions (Table 1) differ by the radius of the CS envelope at the explosion 
   moment: $R_{cs} = 3\times 10^{15}$\,cm in the case A and 
   $R_{cs} = 6\times 10^{15}$\,cm in the case B.
   The Table 1 contains the SN ejected mass, explosion energy, $^{56}$Ni mass, the mass and radius 
   of the CS shell.
 The mass of the lost envelope is $M_{cs} = 8$\msun, another 2\msun\ is presumably lost by the 
   blue supergiant and RSG winds.
  Given 1.4\msun\ of the neutron star, the expected supernova ejecta mass is 2.6\msun.
Adopted supernova parameters are close to those of SN~IIb SN~1993J (Utrobin 1994).   
 
%========================================================
\begin{table}
	\vspace{6mm}
	\centering
	{{\bf Table 1.} Model parameters}\\
	\bigskip	 
	\begin{tabular}{cccccc} 
		\hline	
	 Model   &	$M_{sn}/M_{\odot}$ & $E_{sn}$ (erg) & $M_{ni}/M_{\odot}$ & $M_{cs}/M_{\odot}$ & $R_{cs}$ (cm) \\
	   	\hline	
	 	 \vspace{0.1cm}    
	A    &   2.6  &       $1.3\times10^{51}$ & 0.06 &   8  &    $3\times10^{15}$ \\
	B    &   2.6  &       $1.3\times10^{51}$ &  0.06 &  8  &    $6\times10^{16}$ \\ 
  SN~2006gy	 &   2.6  &     $1.25\times10^{51}$&  0.08 &  8  &    $5.5\times10^{16}$  \\ 	
		\hline
	\end{tabular}
\end{table}
%=======================================================
%
  
 The supernova density distribution is approximated as $\rho = \rho_0/(1 + x^8)$, where 
  $x = v/v_0$ with $\rho_0$ and $v_0$ determined by the ejecta mass and explosion energy.
The CS shell density is assumed to be Gaussian $\rho(r) \propto \exp{(-z^2)}$, with  $z = (r/r_{cs}-1)/\alpha$ and 
  $\alpha = 0.3$ upon the wind background with a moderate density 
   $w = 4\pi r^2\rho = 5\times10^{13}$\gcm.   
The SN/CSM interaction hydrodynamics is treated in a thin shell approximation 
 (Guilliani 1982; Chevalier 1982; Chugai 2001).
 The supernova luminosity is calculated in  the Arnett approximation (Arnett 1980), whereas 
  the luminosity powered by the CS interaction is equal to the instant radiative luminosity of the forward and reverse shocks.

Bolometric light curve combined with the maximum velocity of unshocked ejecta and velocity of 
 the cold dense shell (CDS) formed  in between   
   forward and reverse shocks, are shown in Figures 1 and 2 for A and B model, respectively. 
The difference of the light maximum  epoch and light curve width is the outcome of the 
  different radii of the CS shells. 
 Note the rapid CDS deceleration  at $t \lesssim 100$ day that should be accompanied by 
   the Ralaygh-Taylor instability and the CDS fragmentation. 
The latter effect shortens the diffusion time and thus justifies the omission of a radiation trapping 
in the light curve calculation.

 \section{SN~2006gy: NCSN interacting with lost CE}
 
 The bolometric light curve of SN~2006gy (Smith et al. 2010) is very much similar to that    of the model B, which suggests that SN~2006gy might be the NCSN interacting with the  recently lost CE.
 We rely on the observational bolometric light curve (Jerkstrand et al. 2020) 
  and the maximum expansion velocity $v_{max} = 2900\pm100$\kms\ recovered from 
   \fei\ 7912\AA\ and 8204\AA\  in  
  the spectrum on day 400  (Kawabata et al. 2009; Jerkstrand et al. 2020).
     
 The optimal model of SN~2006gy (Figure 3) describes  the light curve and maximum expansion velocity. 
 The model parameters (Table 1) are comparable ro those of the model B. 
The $^{56}$Ni mass (0.08\msun) is adopted based on the SN~1993J (Utrobin 1996) since 
 the light curve of SN~~2006gy is not sensitive to the amount of $^{56}$Ni. 
The fact that the maximum velocity of unshocked ejecta coincides  with the maximum velocity of the  
  \fei\ line-emitting region strongly suggests that the unshocked ejecta dominates the \fei\ 
    line luminosity.
This conclusion is consistent with the adopted $^{56}$Ni mass that significantly 
  exceeds the iron mass $\approx 0.01$\msun\ in 
  the CS envelope assuming the solar abundance.
  
The proposed scenario for SN~2006gy can be verified via reproducing     
 \fei\ line flux on day 400.
To this end, the most convenient is unblended \fei\ 7912\,\AA\ line of the  $^5$F - $^7$D$^o$ multiplet. 
The transition between levels  $J_1 = 5$, $J_2 = 4$ with excitation potential $E_1 = 0.86$ eV, $E_2 = 2.42$ eV  and the  
spontaneous emission probability  $A_{21} = 168$ s$^{-1}$  (NIST database)
that is high enough to dominate the collisional deexcitation.

In order to calculate the line luminosity we adopt a homologously expanding supernova 
 envelope of the uniform density $\rho_0 = const$.
With the determined mass and energy (Table 1) 
  the boundary velocity is $v_0 = 8950$\kms. 
Almost all the SN ejecta on day 400, therefore, is shocked and merged with the CDS.  
The \fei\ 7912\,\AA\ is emitted entirely by unshocked ejecta with the maximum velocity 
 $v_{max} = 2900$\kms, the mass $M_{ue} = 0.09$\msun.
 The iron mass produced by the 
   $^{56}$Ni decay is assumed mostly, $M(\mbox{Fe}) = 0.06$\msun, to 
   reside in the unshocked ejecta.
 The iron is presumably distributed in the form of clumps 
  with the density exceeding the average density  $\rho(\mbox{Fe}) = \chi \rho_0$
 and the total volume  $V =  M(\mbox{Fe})/(\rho_0\chi)$.

The line emissivity is assumed to be due to only the collisional excitation in the 
  line transiton 
 with the subsequent spontaneous emission  of this line, 
  i.e., collisional and radiative transitions between other levels of the considered  multiplet  
   are neglected.
The lower level is assumed to have Boltzmann population.
The collisional strength for this transition is $\omega_{12} = 2.93(T_e/5000)$  
  (Bautista et al. 2017). 
 The electron temperature is controlled 
by both, the $^{56}$Co radioactive decay and ionizing radiation of the reverse shock, 
so it should exceed the typical value for CCSN without CS interaction at the nebular  
stage ($\sim 5000$\,K).
We consider the  electron temperature as a free parameter in addition to 
 the iron density contrast $\chi$ and \fei\ ionization fraction.
 
 The observed line flux (Kawabata et al. 2009) for the distance of 73.1 Mpc (Smith et al. 2007) 
 suggests the line luminosity $L(7912\mbox{\AA}) = 1.4\times10^{39}$\ergs. 
 In our model this luminosity is reproduced  for 
 $T_e =7500$\,K. $\chi = 10$, and \fei\ ionization fraction of 0.5; 
 these values are sensible.
 One can conclude, therefore, that 
  the proposed scenario for SN~2006gy is qualitatively consistent to the flux of \fei\ lines on day 400.

%================================================================
\begin{figure}
	\centering
	\includegraphics[trim=0 200 0 100,clip, width=\columnwidth]{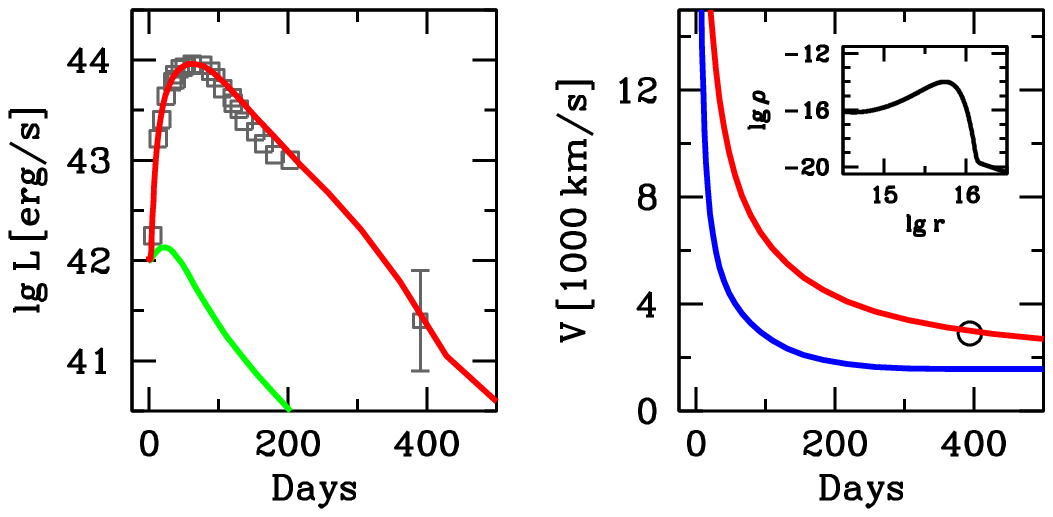}
	\caption{
		{\em Left.} Bolometric light curve of SN~2006gy (Jerkstrand et al. 2020) (squares) with the overplotted model ({\em red}); green line is the luminosity of the exploded NCSN. 
		{\em Right.} CDS velocity ({\em blue}) and maximum velocity of unshocked ejecta 
			({\em red}); maximum velocity of the line-emitting gas on day 400 is shown by 
				the circle. Inset shows the CS density.
				}
	\label{fig:06gy}
\end{figure}
%==================================================================

\section{Conclusions}
	
The paper presents the solution to the problem of synchronization between 
  the presupernova RSG loss of the hydrogen envelope and supernova explosion followed by 
  the subsequent interaction  with the lost massive envelope.
The central to the proposed mechanism is the conjecture on a moderate RSG expansion  
  during the final 10-20 years before the core collapse.
 The  RSG  expansion results in the finite probabily for a close  
 component  to plunge into the RSG envelope thus turning on the CE regime at the right time. 
 The idea of the RSG expansion is based on the phenomenon of the presupernova DCS 
  that forms due to the vigorous mass loss during the final 10-20 years before the explosion.

The successful modelling for the SN~2006gy light curve and the expansion velocity 
  of the \fei\ line-emitting zone combined with the explanation of \fei\ 7912\,\AA\ luminosity on day 400  suggests 
  the SN~2006gy origin from the NCSN explosion inside the recently lost CE.

\vspace{1cm}  
\centerline{\bf Acknowledgements}
\bigskip  

I am gratefull to Lev Yungelson, Maxim Barkov and Ludmila Mashonkina for useful discussions.

\clearpage

\centerline{\bf References}
\bigskip

\noindent
 Arnett W. D. , Astrophys. J. {\bf 237}, 541 (1980)\\
\medskip
 Barkov M. V., Komissarov S. S., Mon. Not. R. Astron. Soc. {\bf 415}, 944 (2011)\\
\medskip
Bautista M. A., Lind K., Bergemann M., Astron. Astrophys. {\bf 606}, A127 (2017)\\
\medskip
Chugai N. N., Mon. Not. R. Astron. Soc. {\bf 508}, 6023 (2021)\\
\medskip
Chugai N. N., Chevalier R. A.,  Astrophys. J. {\bf 641}, 1051 (2006)\\       
\medskip
Chugai) N. N., Mon. Not. R. Astron. Soc. {\bf 326}, 1448 (2001)\\
\medskip
Chugai N. N., Danziger I. J.,  Mon. Not. R. Astron. Soc. {\bf 268}, 173 (1994)\\
\medskip
Chevalier R. A., Astrophys. J. {\bf 752}, 2 (2012)\\
\medskip
Chevalier R. A., Astrophys. J. {\bf 259}, 302 (1982)\\
\medskip
Duch\.{e}ne G., Kraus A., Ann. Rev. Astron. Astrophys. {\bf 51}, 269 (2013)\\  
\medskip
Giuliani J. L.,  Astrophys. J. {\bf 256}, 624 (1982)\\  
\medskip
Han Z. W., Ge H. W., Chen X. F., Chen H. L., Res. Astron. Astrophys. {\bf 20}, 161 (2020)\\
\medskip
Ivanova N., Justham S., Chen X. et al.,  Astron. Astrophys. Rev {\bf 21}, id 59  (2013)\\
\medskip
Jerkstrand A., Smartt S. J., Inserra C. et al.,  Astrophys. J. {\bf 835}, 13 (2017)\\
\medskip
Kawabata K. S., Tanaka M., Maeda K., et al., Astrophys. J. {\bf 697}, 747 (2009)\\
\medskip
Khazov D., Yaron O., Gal-Yam A. et al., Astrophys. J. {\bf 818}, 3 (2016)\\  
\medskip
Kruckow M. U., Tauris T. M, Langer N. et al., Astron. Astrophys. {\bf 596}, A58 (2016)\\
\medskip
Li W., Leaman J., Chornock R. et al., Mon. Not. R. Astron. Soc. {\bf 412}, 1441 (2011)\\
\medskip
 Miller A. A., Silverman J. M., Butler N. R. et al., Mon. Not. R. Astron. Soc. {\bf 404}, 305 (2010)\\
\medskip
Ofek E. O., Cameron P. B., Kasliwal M. M. et al., Astrophys. J. {\bf 659}, L13  (2007)\\
\medskip
 \"{O}pik E., Publ. Obs. Astron. Univ. Tartu  {\bf 25},   No.6, 1 (1924) \\
\medskip
Podsiadlowski Ph., Joss P. C., Hsu J. J. L.,  Astrophys. J. {\bf 391}, 246  (1992)\\
\medskip
Popova E. I., Tutukov F. V., Yungelson) L. R., Astrophys. Space Sci. {\bf 88},
55 (1982)\\
\medskip
Shiode J. H., Quataert E., Astrophys. J. {\bf 780}, 96 (2014)\\
\medskip
Smith1 N., Chornock1  R., Silverman J. M. et al., Astrophys. J. {\bf 709}, 856  (2010)\\
\medskip
Smith N., Li W., Foley R. J. et al.,  Astrophys. J. {\bf 666}, 1116 (2007)\\
\medskip
Soderberg A. M, Nakar E., Berger E., Kulkarni S. R., Astrophys. J. {\bf 638}, 930 (2006)\\
\medskip
Utrobin V. P., Astron. Astrophys.  {\bf 281}, L89  (1994)\\
\medskip
Utrobin V. P., Astron. Astrophys.  {\bf 306}, 219  (1996)\\
\medskip
Vereshchagin S., Tutukov A., Yungelson L., et al., Astrophys. Space Sci. {\bf 142}, 245 (1982)\\
\medskip
Woosley S. E., Blinnikov S. I., Heger A., Nature {\bf 450}, 390 (2007)\\
\medskip
Woosley S. E., Heger A., T. A. Weaver N. A., Review Mod. Phys. {\bf 74}, 1015 (2002)\\
\medskip
Yaron O., Perley  D. A., Gal-Yam  A. et al., Nature Physics, {\bf 13},  510 (2017)\\.

\end{document}